\documentclass[prb,twocolumn,superscriptaddress,showpacs,epsf,preprintnumbers]{revtex4}

\usepackage{ulem} 
\usepackage{epsfig}
\usepackage{color}

\usepackage{graphicx}
\usepackage{latexsym}
\usepackage{amsmath}
\usepackage{amssymb}
\usepackage{amsfonts}

\newcommand{\gapprox}{{\scriptscriptstyle\stackrel{>}{\sim}}}

\newif\ifgraph

\graphtrue                   %

\begin{document}

\title{Crossover between different regimes of inhomogeneous superconductivity \\
in planar superconductor--ferromagnet hybrids}

\author{A.Yu. Aladyshkin}
\affiliation{INPAC -- Institute for Nanoscale Physics and
Chemistry, K.U. Leuven, Celestijnenlaan 200D, B--3001 Leuven,
Belgium} \affiliation{Institute for Physics of Microstructures
RAS, 603950, Nizhny Novgorod, GSP-105, Russia}
\author{J. Fritzsche} \affiliation{INPAC -- Institute for Nanoscale Physics
and Chemistry, K.U. Leuven, Celestijnenlaan 200D, B--3001 Leuven,
Belgium}
\author{R. Werner}
\affiliation{Physikalisches Institut -- Experimentalphysik II,
Universit\"{a}t T\"{u}bingen, Auf der Morgenstelle 14, 72076
T\"{u}bingen, Germany}
\author{R.B.G.~Kramer} \affiliation{INPAC --
Institute for Nanoscale Physics and Chemistry, K.U. Leuven,
Celestijnenlaan 200D, B--3001 Leuven, Belgium}
\affiliation{Institut N\'eel, CNRS--Universit\'e Joseph Fourier,
BP 166, 38042 Grenoble Cedex 9, France}
\author{S.~Gu\'{e}non}
\affiliation{Physikalisches Institut -- Experimentalphysik II,
Universit\"{a}t T\"{u}bingen, Auf der Morgenstelle 14, 72076
T\"{u}bingen, Germany}
\author{R. Kleiner}
\affiliation{Physikalisches Institut -- Experimentalphysik II,
Universit\"{a}t T\"{u}bingen, Auf der Morgenstelle 14, 72076
T\"{u}bingen, Germany}
\author{D. Koelle}
\affiliation{Physikalisches Institut -- Experimentalphysik II,
Universit\"{a}t T\"{u}bingen, Auf der Morgenstelle 14, 72076
T\"{u}bingen, Germany}
\author{V.V. Moshchalkov} \affiliation{INPAC
-- Institute for Nanoscale Physics and Chemistry, K.U. Leuven,
Celestijnenlaan 200D, B--3001 Leuven, Belgium}

\date{\today}
\begin{abstract}
We studied experimentally the effect of a stripe-like domain
structure in a ferromagnetic BaFe$_{12}$O$_{19}$ substrate on the
magnetoresistance of a superconducting Pb microbridge. The system
was designed in such a way that the bridge is oriented
perpendicular to the domain walls. It is demonstrated that
depending on the ratio between the amplitude of the nonuniform
magnetic field $B_0$, induced by the ferromagnet, and the upper
critical field $H_{c2}$ of the superconducting material, the
regions of the reverse-domain superconductivity in the $H-T$ plane
can be isolated or can overlap ($H$ is the external magnetic
field, $T$ is temperature). The latter case corresponds to the
condition $B_0/H_{c2}<1$ and results in the formation of
superconductivity above the magnetic domains of both polarities.
We discovered the regime of edge-assisted reverse-domain
superconductivity, corresponding to localized superconductivity
near the edges of the bridge above the compensated magnetic
domains. Direct verification of the formation of inhomogeneous
superconducting states and external-field-controlled switching
between normal state and inhomogeneous superconductivity were
obtained by low-temperature scanning laser microscopy.
\end{abstract}



\maketitle

\section{Introduction}

Recent advances in fabrication technology have made it possible to
realize superconductor--ferromagnet (S/F) hybrid structures with a
controlled arrangement of the ferromagnetic layers/elements. These
flux--coupled and exchange--coupled S/F
hybrids\cite{Buzdin-RMP-05,Lyuksyutov-AdvPhys-05,Velez-JMMM-08,Aladyshkin-SuST-09}
are of fundamental interest for investigations of the nontrivial
interaction between superconductivity and a nonuniform
distribution of a local magnetic field (or magnetization).
Furthermore, S/F hybrids seem to be potential candidates for the
development of tunable elements of superconducting
electronics.\cite{Aladyshkin-SuST-09} In this paper we focus only
on the planar S/F structures, consisting of a low-$T_c$
superconducting film and a ferromagnetic layer with a domain
structure with a dominant magnetostatic interaction between
superconducting and ferromagnetic films.
\cite{Gillijns-PRL-05,Gillijns-PRB-07,
Aladyshkin-JAP-10,Yang-Nature-04,Fritzsche-PRL-06,Yang-APL-06,Yang-PRB-06,Aladyshkin-APL-09,Aladyshkin-PhysC-10,
Aladyshkin-APL-10,Zhu-PRL-08,Zhu-PRB-10,Haindl-SuST-08,Fritzsche-PRB-09,VlaskoVlasov-PRB-08-a,VlaskoVlasov-PRB-08-b,VlaskoVlasov-PRB-10,Bell-PRB-06,
Rakshit-PRB-08-a,Rakshit-PRB-08-b,Belkin-APL-08,Belkin-PRB-08,Karapetrov-PRB-09,Belkin-APL-10,Visani-PRB-10,Tamegai-SuST-11}

\begin{figure}[t!]
    \begin{center}
    \ifgraph\includegraphics[width=8.5cm]{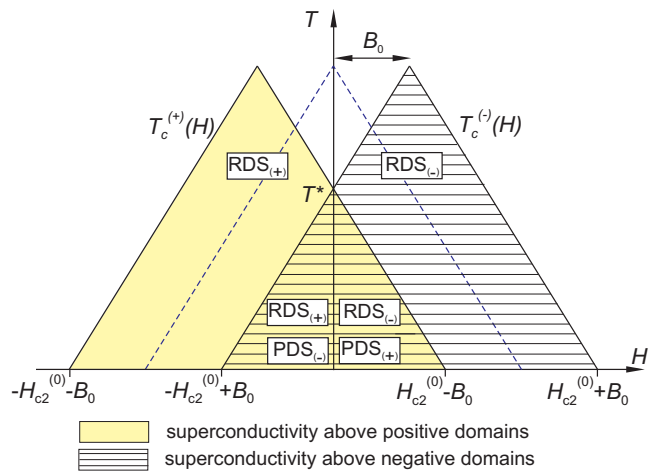}\fi
    \end{center}
    \caption{(color online) Transformation of the phase diagram
    ``external magnetic field ($H$) -- temperature $(T)$"  in the
    presence of a nonuniform magnetic field
    ($B_0$ is the amplitude of the $z-$component of the stray field).
    The dashed line is plotted according to Eq.~(\ref{Hc2}).
    The filled (shaded) area corresponds to inhomogeneous
    superconductivity
    above positive and negative magnetic domains, respectively, see
    Eqs.~(\ref{Biased-Hc2a})--(\ref{Biased-Hc2b}). Here we use the
    notations: RDS$_{(\pm)}$ -- reverse-domain superconductivity
    localized above positive ($+$) domains at $H<0$ and above negative ($-$) domains at $H>0$;
    PDS$_{(\pm)}$ -- parallel-domain superconductivity
    localized above positive ($+$) domains at $H>0$ and above negative ($-$) domains at
    $H<0$.
    } \label{Fig-PBs}
    \end{figure}

The nonuniform component of the magnetic field, induced by a
ferromagnet, can modify the conditions for the appearance of
superconductivity in thin superconducting films due to the effect
of field compensation. The formation of localized
superconductivity in the areas of compensated magnetic
field\cite{Pannetier-95,Lange-PRL-03,Aladyshkin-PRB-03} results in
an exotic dependence of the superconducting critical temperature
$T_c$ on the external magnetic field $H$, applied perpendicular to
the thin film plane. This phase transition line $T_c$ vs. $H$ for
planar S/F hybrids becomes non-monotonous
\cite{Gillijns-PRL-05,Gillijns-PRB-07,Aladyshkin-JAP-10,Yang-Nature-04,
Fritzsche-PRL-06,Yang-APL-06,Yang-PRB-06,Aladyshkin-APL-09,Aladyshkin-PhysC-10,
Aladyshkin-APL-10,Zhu-PRL-08,Zhu-PRB-10,Haindl-SuST-08,Lange-PRL-03,Aladyshkin-PRB-03,Buzdin-PRB-03,Aladyshkin-PRB-06}
and thus, it differs significantly from the standard linear
dependence of the upper critical field $H_{c2}$ on temperature
$T$, which can be written as
    \begin{eqnarray}
    1-\frac{T_c}{T_{c0}} = \frac{|H|}{H_{c2}^{(0)}}.
    \label{Hc2}
    \end{eqnarray}
Here $T_{c0}$ is the critical temperature of the superconducting
transition at zero magnetic field,
$H_{c2}^{(0)}=\Phi_0/(2\pi\xi_0^2)$ and $\xi_0$ are upper critical
field and coherence length at $T=0$, respectively, and
\mbox{$\Phi_0=\pi\hbar c/e$} is the magnetic flux quantum. Such
dependence given by Eq.~(\ref{Hc2}) is inherent for plain
superconducting films in a uniform magnetic field $H$ and shown by
the dashed line in Fig.~\ref{Fig-PBs}.

Considering qualitatively the effect of an inhomogeneous magnetic
field, which varies periodically from $+B_0$ (above positive
magnetic domains) to $-B_0$ (above negative domains) and remains
constant inside the domains, one can expect two different phase
transition lines
    \begin{eqnarray}
    1-\frac{T_{c}^{(+)}}{T_{c0}} =
    \frac{|H+B_0|}{H_{c2}^{(0)}},\quad \mbox{(positive domains)}, \label{Biased-Hc2a}\\
    1-\frac{T_{c}^{(-)}}{T_{c0}} =
    \frac{|H-B_0|}{H_{c2}^{(0)}},\quad \mbox{(negative domains)}.
    \label{Biased-Hc2b}
    \end{eqnarray}
These two different transition lines correspond to the formation
of superconductivity in the areas where the perpendicular
$z-$component of the nonuniform magnetic field is positive,
Eq.~(\ref{Biased-Hc2a}), or negative, Eq.~(\ref{Biased-Hc2b}). The
phase diagram $H-T$, composed according to
Eqs.~(\ref{Biased-Hc2a})--(\ref{Biased-Hc2b}), is shown in
Fig.~\ref{Fig-PBs}. An inhomogeneous superconducting state trapped
only within the areas above the domains of the opposite polarity
with respect to the sign of $H$, is commonly referred to as
reverse-domain superconductivity
(RDS).\cite{Yang-Nature-04,Aladyshkin-PRB-06,Fritzsche-PRL-06,Yang-APL-06,Yang-PRB-06,Aladyshkin-APL-09}
In order to guarantee the formation of such RDS exclusively above
either positive domains (at $H<0$) or above negative domains (at
$H>0$), one should satisfy $B_0/H_{c2}>1$ (i.e., at high
temperatures or/and large $B_0$ values). Such separated regions of
RDS above the domains of different signs were observed
experimentally\cite{Yang-APL-06,Yang-PRB-06,Aladyshkin-APL-09,Aladyshkin-PhysC-10,Aladyshkin-APL-10,Fritzsche-PRB-09}
for Pb/BaFe$_{12}$O$_{19}$, Nb/BaFe$_{12}$O$_{19}$ and
Al/BaFe$_{12}$O$_{19}$ hybrid structures. Upon decreasing $T$
and/or $B_0$, inhomogeneous superconductivity above magnetic
domains of both polarities (i.e. parallel and antiparallel) can
coexist (cf. Fig.~\ref{Fig-PBs}), since both criteria for the
formation of inhomogeneous superconductivity above both magnetic
domains, $|H-B_0|<H_{c2}$ and $|H+B_0|<H_{c2}$, can be fulfilled
simultaneously --- provided
    \begin{eqnarray}
    -\left(1-\frac{B_0}{H_{c2}}\right)\, < \,\frac{H}{H_{c2}}
    \,<\, \left(1-\frac{B_0}{H_{c2}}\right).
    \label{H-range2}
    \end{eqnarray}
The threshold temperature $T^*$ of the crossover from simple RDS
to a complex state consisting of RDS in the compensated regions
and parallel domain superconductivity (PDS) in the regions with
enhanced magnetic field corresponds to the intersection point of
the dependencies $T_c^{(+)}(H)$ and $T_c^{(-)}(H)$ at \mbox{$H=0$}
and can be estimated as $T^*=T_{c0}(1-B_0/H_{c2})$. Such an
inhomogeneous superconducting state, potentially observed in the
$H$ range described by Eq.~(\ref{H-range2}) and characterized by a
development of superconductivity above magnetic domains of both
polarities, can be also called complete superconductivity (CS).


For superconducting samples of finite lateral dimensions one can
expect the appearance of surface
superconductivity,\cite{Saint-James-69,Tinkham-96} i.e.
superconductivity localized near the sample edges even in the
presence of a nonuniform magnetic field. In the first
approximation, the phase transition line for such anticipated
edge-assisted (EA) superconductivity can be described by the
shifted $H_{c3}$ dependencies
    \begin{eqnarray}
    1-\frac{T_c^{(\pm)}}{T_{c0}} = 0.59\,\frac{|H\pm
    B_0|}{H_{c2}^{(0)}},
    \label{Biased-Hc3}
    \end{eqnarray}
where the signs $+/-$ correspond to edge-assisted
superconductivity above positive/negative magnetic domains,
respectively. This means that superconductivity near the sample's
edges will survive until the local field exceeds the critical
field of surface superconductivity $H_{c3}=1.69\,H_{c2}$.
Furthermore, the presence of the domain walls stimulates the
formation of domain-wall superconductivity (DWS) for moderate
fields, $|H|\le B_0$.\cite{Buzdin-PRB-03,Aladyshkin-PRB-03} It has
been shown that for thin superconducting films (in the $(x,y)$
plane) in a perpendicular magnetic field $B_z= H + b_z(x)$ with a
step-like component $b_z(x)=B_0\,{\rm sign}(x)$ induced by a
domain wall, nucleation of the superconducting order parameter
along the domain wall (at $x=0$) becomes possible below the phase
transition line\cite{Comment-2}

    \begin{eqnarray}
    \nonumber 1-\frac{T^{DWS}_c}{T_{c0}} \simeq
    \frac{B_0}{H_{c2}^{(0)}}
    \times \\
    \left\{0.59 - 0.70\left(\frac{H}{B_0}\right)^2 + 0.09\left(\frac{H}{B_0}\right)^4
    \right\}.
    \label{Aladyshkin-DWS}
    \end{eqnarray}
Thus, the set of possible nonuniform superconducting solutions in
flux--coupled S/F hybrids has to include the following states:
domain-wall superconductivity (DWS), ``bulk" RDS and PDS above
either positive or negative domains, edge-assisted RDS and PDS
above either positive or negative domains, and CS above the
domains of both polarities.

In this paper we present an experimental study of the
temperature-- and field--induced crossover between the different
regimes of bulk and localized superconductivity in S/F hybrid
structures, consisting of a thin superconducting Pb film on top of
a bulk ferromagnetic BaFe$_{12}$O$_{19}$ single crystal with a
well defined stripe-like domain structure. In order to exclude
percolation effects and electrical shunting arising from different
superconducting states, we prepared the hybrid S/F structure such
that the domain walls are oriented {\it across} the
superconducting microbridge. As a result, the current density is
distributed over the entire cross-section of the microbridge, what
allows us to detect variations of the voltage drop associated with
the appearance/destruction of inhomogeneous superconductivity of
various types.



    \begin{figure}[b!]
    \begin{center}
    \includegraphics[width=70mm]{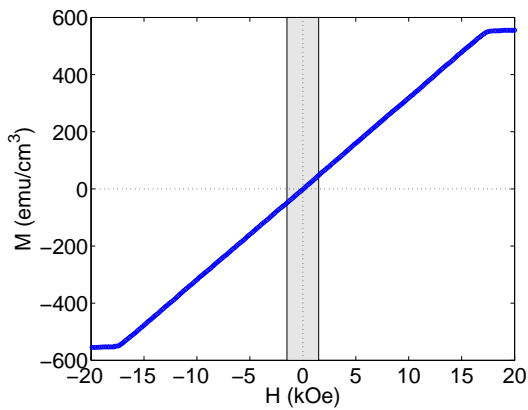}
    \end{center}
    \caption{(color online) Magnetization curve $M$ vs. $H$,
    obtained with a vibrating
    sample magnetometer at $T=5$~K.
    The shaded area indicates the range of $H$ in our experiment.}
    \label{Fig-VSM-magnetometry}
    \end{figure}

\section{Magnetic properties of the ferromagnetic substrate}

For the creation of a static nonuniform magnetic field with a
well-defined domain structure, we used bulk ferromagnetic crystals
BaFe$_{12}$O$_{19}$ (BFO). When cut along the proper
crystallographic direction, these BFO crystals exhibit a
stripe-type domain structure with dominant in-plane magnetization
and a relatively small out-of-plane component of
magnetization.\cite{Aladyshkin-APL-09,Aladyshkin-PhysC-10,Aladyshkin-APL-10,Fritzsche-PRB-09}

Measurements with a vibrating sample magnetometer revealed that at
low temperatures the magnetization of the used crystal depends
almost linearly on the applied perpendicular magnetic field, and
that it saturates at $H\simeq 17$~kOe (see
Fig.~\ref{Fig-VSM-magnetometry}). This means that external
magnetic field $|H|\le 1.5$~kOe, which corresponds to the range of
the $H$-sweeps in our measurements, can only be of minor influence
on the domain structure, since the variation of magnetization of
the substrate is expected to be not more than 9\% of the saturated
magnetization.

    \begin{figure}[t!]
    \begin{centering}
    \includegraphics[height=3.85cm]{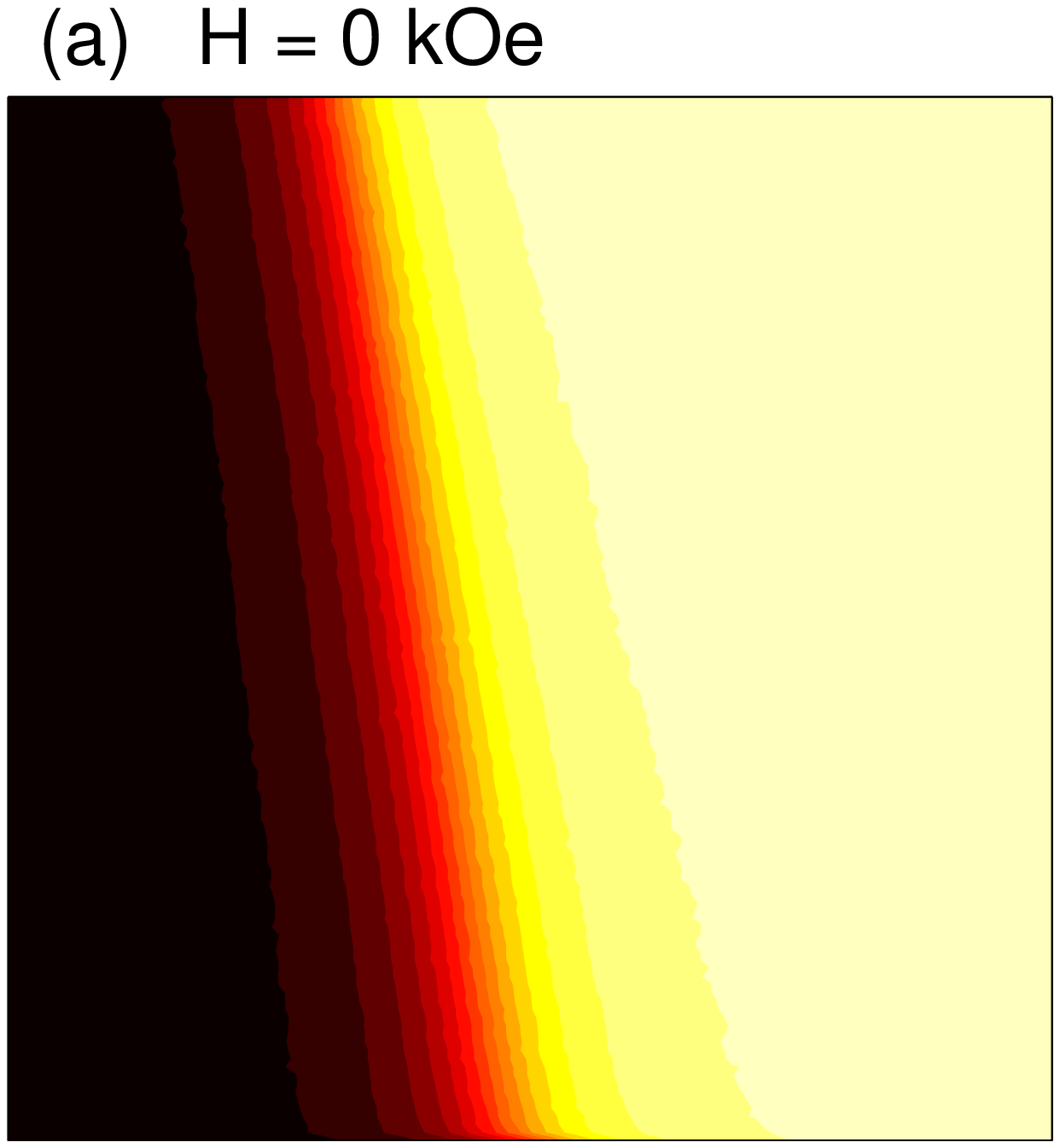}
    \includegraphics[height=3.85cm]{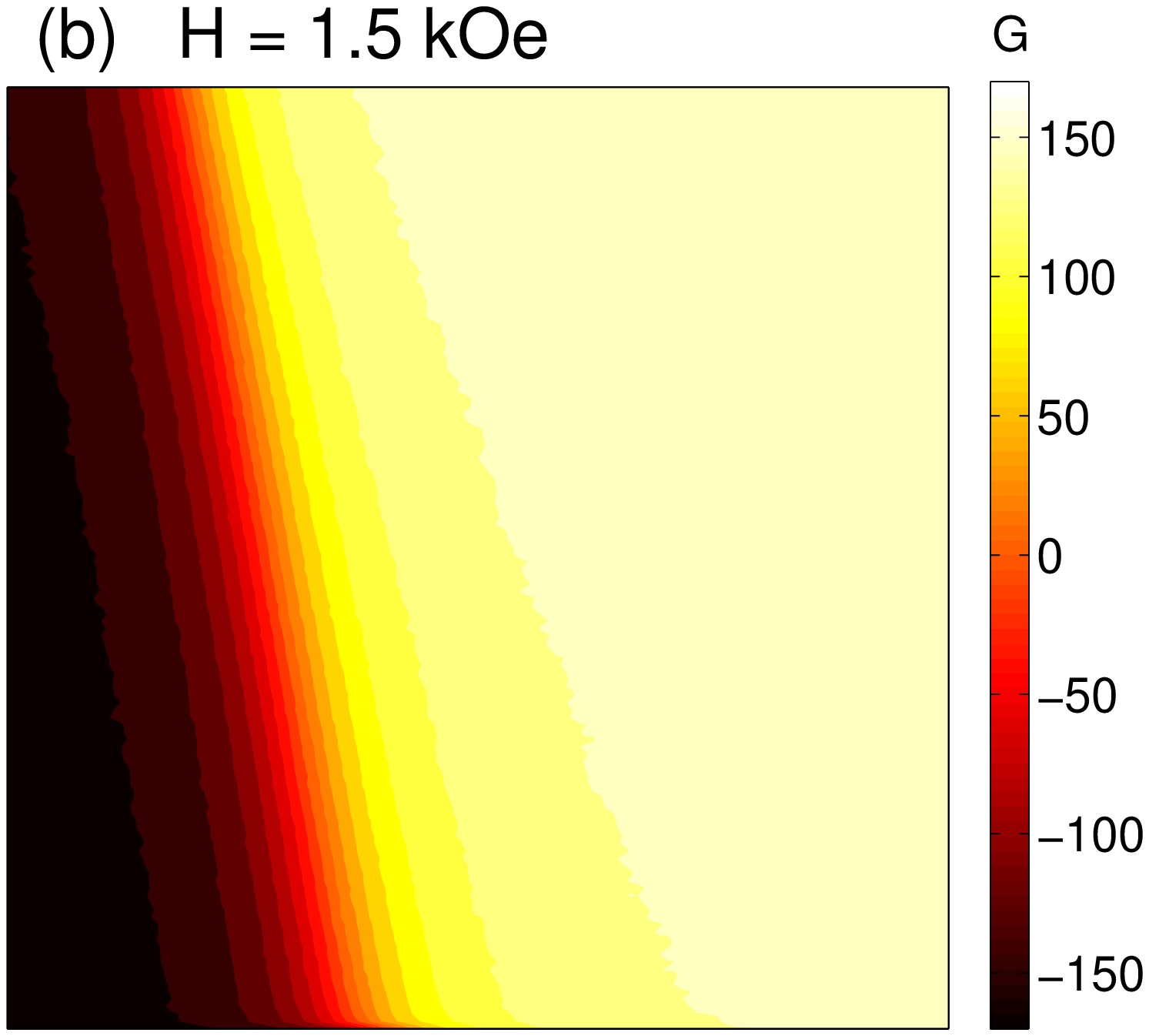}
    \includegraphics[width=7.5cm] {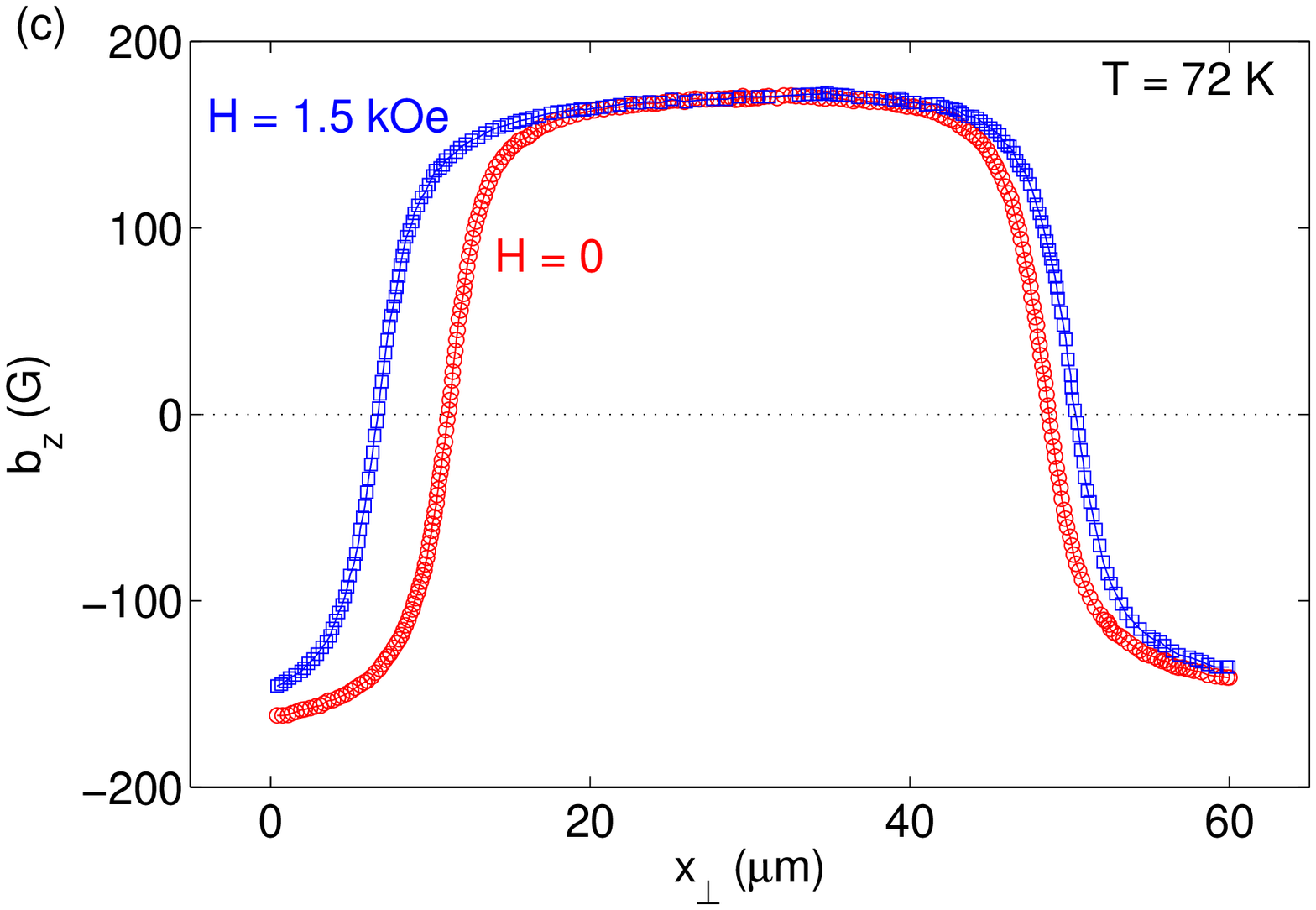}
    \end{centering}
    \caption{(Color online) (a)--(b) Two-dimensional distribution of
    the $z-$component of the nonuniform magnetic field
    $b_z(x,y)=B_z(x,y)-H$ measured
    near a domain wall by a scanning Hall probe
    microscope at the distance 400 nm above the ferromagnetic crystal
    at the external magnetic field $H=0$ (a) and $H=1.5$~kOe (b) at $T=72$ K.
    The scanning area is \mbox{35$\,\mu$m $\times$ 35~$\mu$m}.\newline
    (c) One-dimensional profile of the nonuniform magnetic field
    $b_z(x_{\perp})=B_z(x_{\perp})-H$ along
    the direction perpendicular to two domain walls,
    measured by a scanning Hall probe microscope for the same
    conditions: $H=0$ (red circles) and $H=1.5$~kOe (blue squares).
    } \label{Fig-SHPM-magnetometry}
    \end{figure}

The spatial two-dimensional (2D) distribution of the perpendicular
$z-$component of the magnetic field $b_z$, induced by the laminar
domain structure, was imaged with a scanning Hall probe
microscope\cite{Bending-99} (Fig.~\ref{Fig-SHPM-magnetometry}). By
analyzing the 2D patterns of $b_z(x,y)$, which corresponds to the
difference between the locally measured field $B_z(x,y)$ of the
crystals and the external magnetic field $H$ [panels (a)--(b)], we
come to the following conclusions:

(i) the domain walls remain rectilinear even in the presence of an
external field;

(ii) the effect of the external field is limited mainly to a small
displacement of the domain walls as $H$ varies, leading to a
broadening of the positive magnetic domains at $H>0$ at the
expense of the negative magnetic domains, and vice versa.

The 1D profile of the stray field $b_z$ along a line perpendicular
to the domain walls is shown in
Fig.~\ref{Fig-SHPM-magnetometry}(c). We find that an external
field of the order of 1 kOe shifts the points $b_z\simeq 0$ at
maximum 3$\,\mu$m, which is much less than the equilibrium domain
width of 30$\,\mu$m, without substantial changes in both the shape
of the domain wall and the amplitude of the built-in magnetic
field. This allows us to consider the field pattern as almost
independent on~$H$.

    \begin{figure}[t!]
    \begin{center}
    \ifgraph\includegraphics[width=8.2cm]{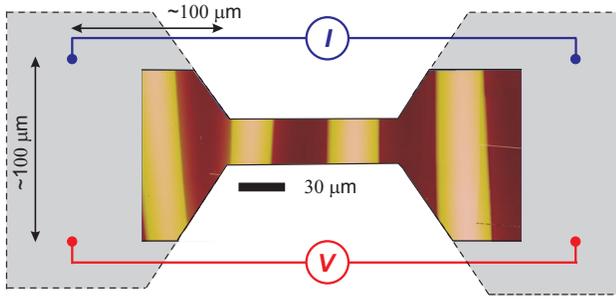}\fi
    \end{center}
    \caption{(color online) Combined MFM and
    optical image of the short Pb thin film bridge fabricated on top of
    the BFO single crystal. The alternating
    dark and bright stripes, corresponding to the magnetic domains with
    different orientation of the magnetic moment, are shown only within
    the superconducting bridge.
    All elements of the electrical circuit as well as the geometry of the contact
    pads (grey areas) are shown schematically.} \label{Fig-System}
    \end{figure}

\section{Superconducting properties of the S/F bilayer}

{\it Sample preparation.} After polishing the cut-surfaces of the
BFO crystals, we prepared lithographically an array of metallic Au
markers ($2\times 2~\mu$m in size) on top of the ferromagnetic
template. The location of the domain walls with respect to the
periodically positioned Au markers was determined with a magnetic
force microscope (MFM) at room temperature. A thin insulating Ge
layer (4 nm thick) was evaporated in order to prevent exchange
interaction and proximity effect between the superconducting and
ferromagnetic layers. Finally, two Pb bridges oriented {\it
across} the domain walls were fabricated by means of e-beam
lithography, molecular beam epitaxy and lift-off technique in a
single run on the same substrate. These superconducting bridges
have the same width (30$~\mu$m) and thickness (40 nm) and differ
only by their lengths: 100 $\mu$m for the short bridge and
700~$\mu$m for the long bridge, resulting in different numbers of
domain walls (4 and 24, respectively) in the narrow part of the
bridges. A combination of an optical and an MFM image of the short
Pb bridge is presented in Fig.~\ref{Fig-System}.

\vspace{0.2cm}

{\it Magnetoresistive curves $R(H)$.} Measurements of the dc
electrical resistance $R$ of both the short and the long
superconducting Pb bridges as a function of temperature $T$, $H$
(applied perpendicularly to the plane of the structures) and the
bias current $I$, were carried out in a commercial Oxford
Instruments cryostat using a conventional four-terminal
configuration.

    \begin{figure}[bh!]
    \includegraphics[width=8.0cm]{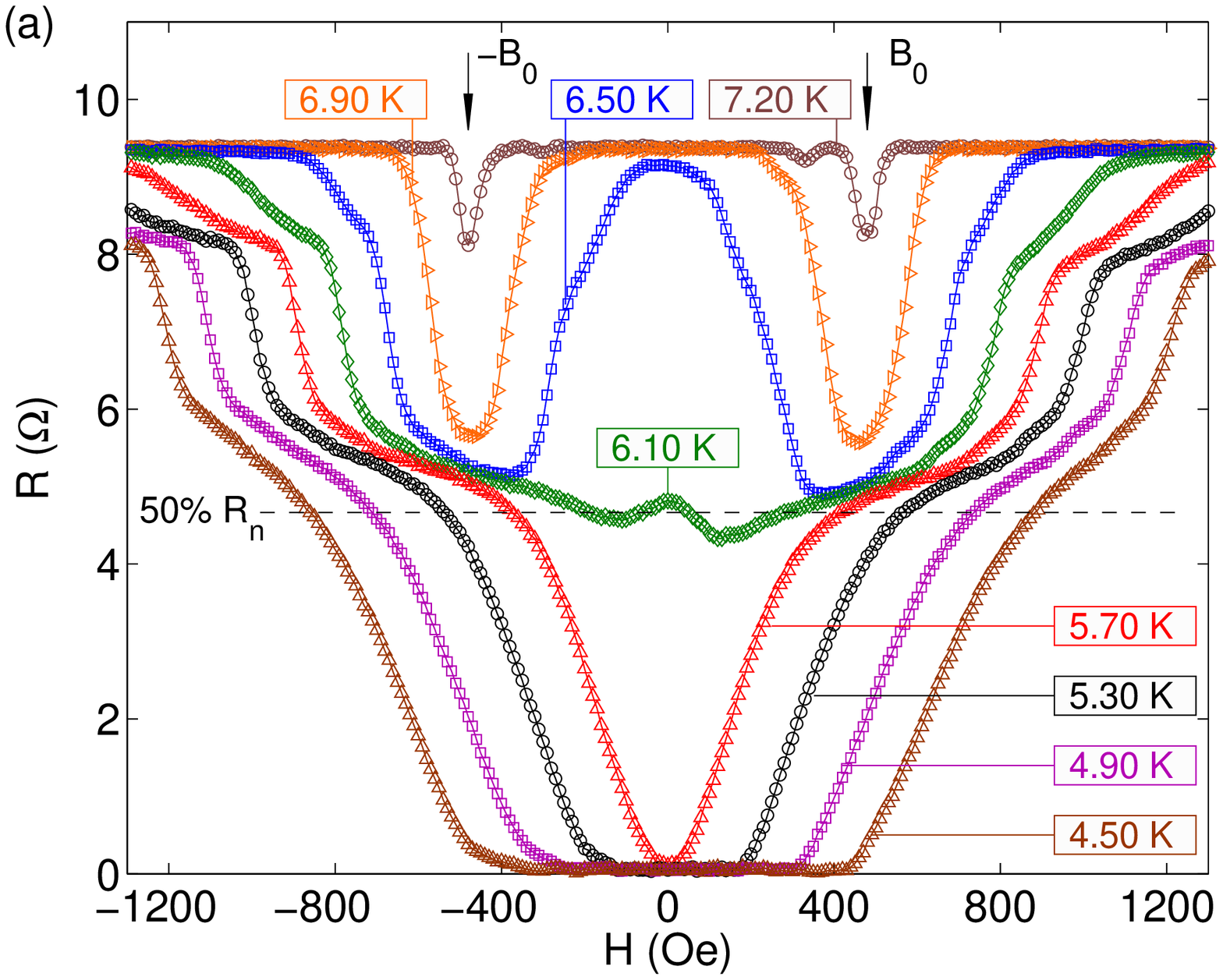}
    \includegraphics[width=8.0cm]{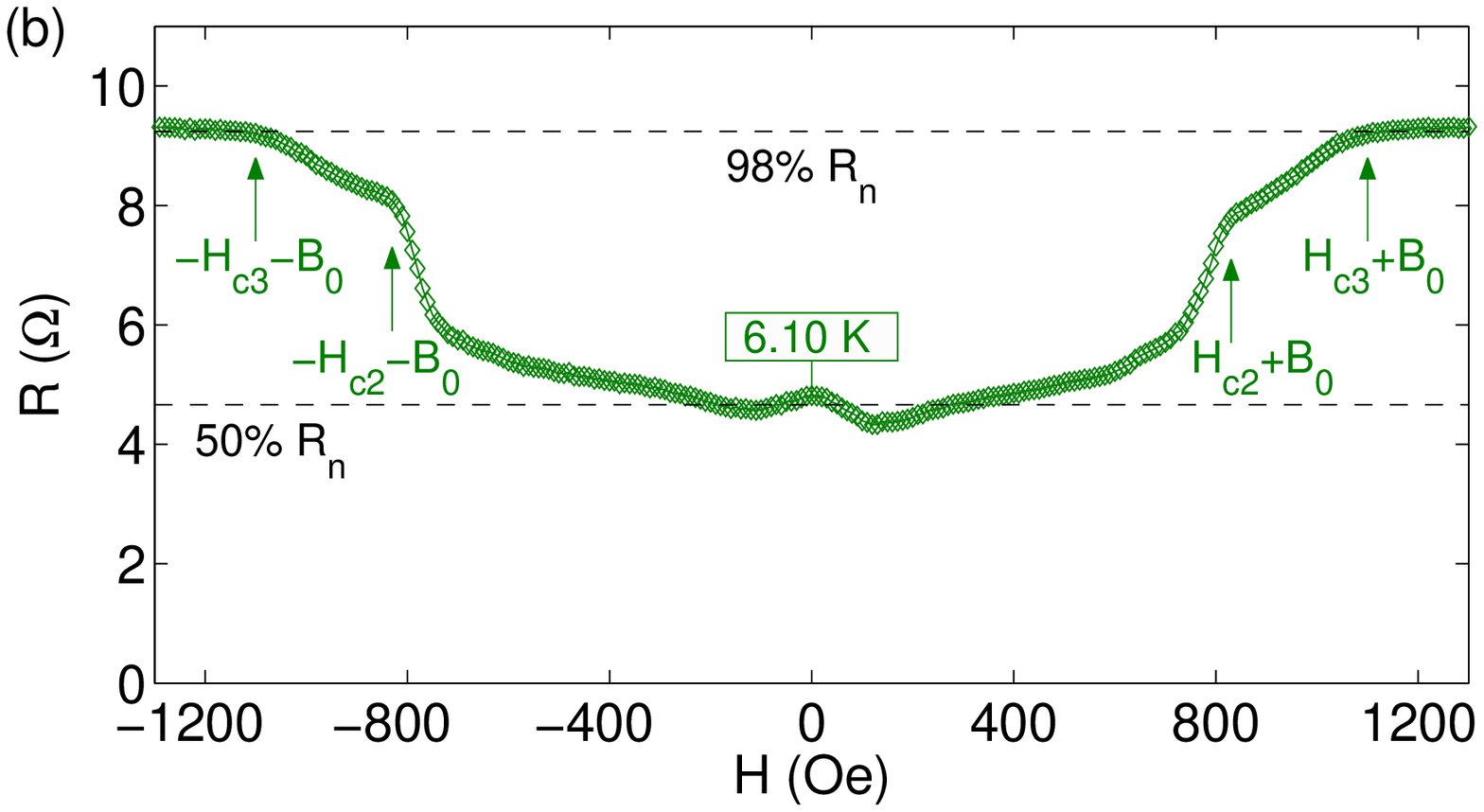}
    \caption{(Color online) (a) Typical dependencies $R$ vs. $H$ taken for the long Pb microbridge at
    a fixed bias current $I=100~\mu$A and
    at different temperatures $T$. \newline
    (b) A single resistive $R(H)$ curve at 6.10~K and the position of the critical fields $|H_{c2}+B_0|$ and
    $|H_{c3}+B_0|$, marked by arrows: $B_0\simeq 480$~Oe, $H_{c2}\simeq 350~$Oe, $H_{c3}\simeq 620$~Oe.
    Two dashed horizontal lines show the levels 0.50$\,R_n$ and 0.98$\,R_n$, where $R_n\approx 9.55\,\Omega$.}
    \label{Fig-resistance}
    \end{figure}


    \begin{figure*}[t!]
    \begin{centering}
    \includegraphics[width=16cm]{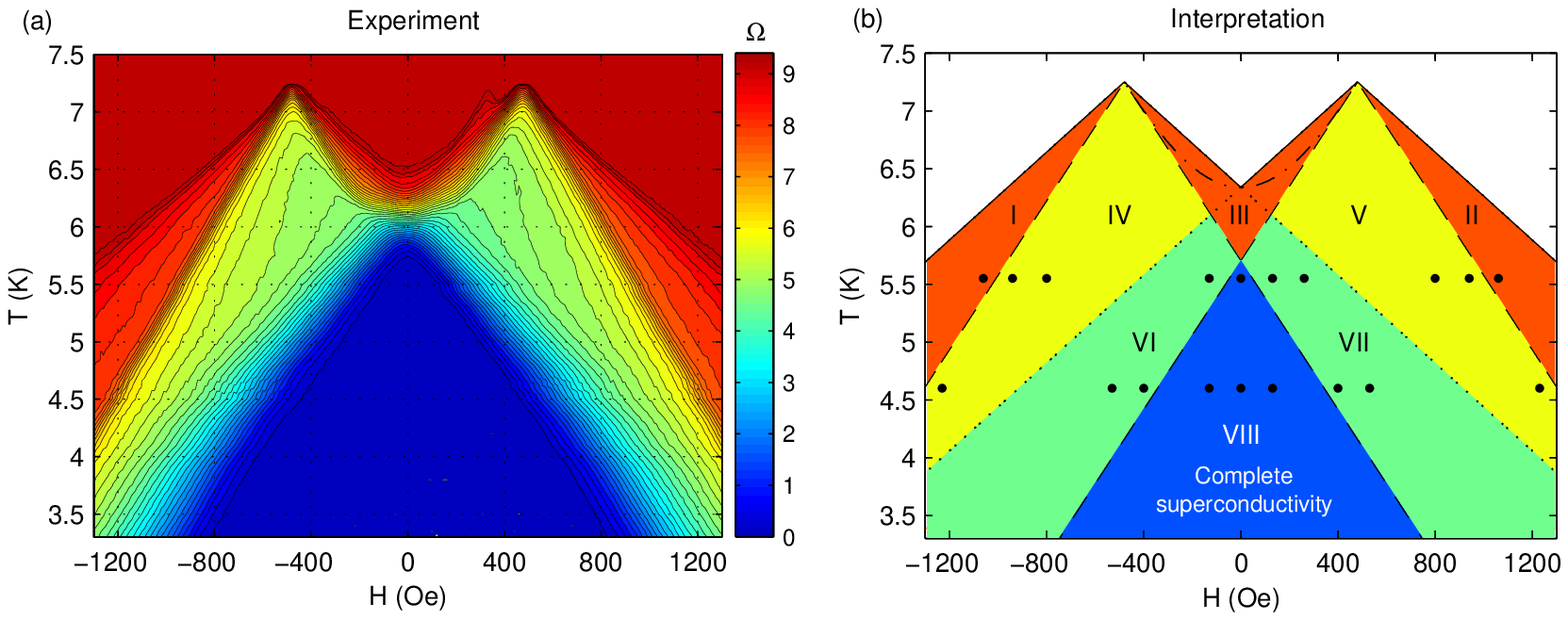}
    \includegraphics[width=18cm]{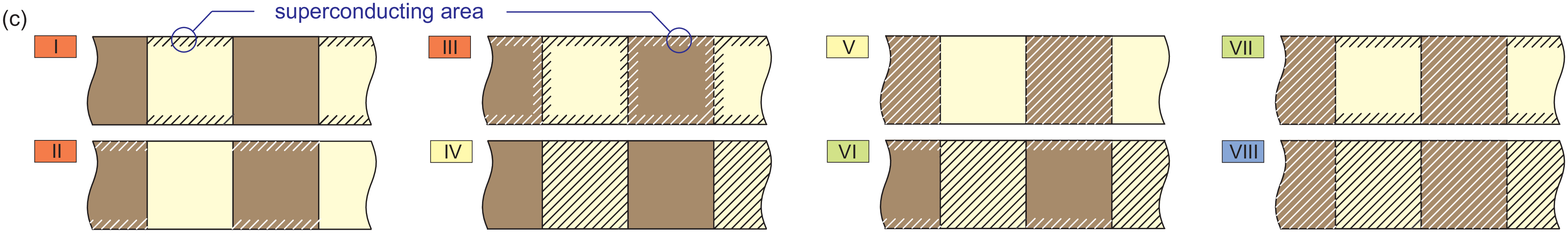}
    \end{centering}
    \caption{(Color online) (a) Dc resistance $R$ of the
    superconducting Pb bridge fabricated on top of the BFO crystal as a
    function of the external magnetic
    field $H$ and temperature $T$ at $I=100~\mu$A. \newline
    (b) Interpretation of the measured $R(H,T)$ dependence:
    solid lines describe the appearance of the edge-assisted reverse-domain superconductivity (EA-RDS),
    Eq.~(\ref{Biased-Hc3});
    dashed lines correspond to the appearance of reverse-domain superconductivity (RDS)
    and parallel-domain superconductivity (PDS),
    Eqs.~(\ref{Biased-Hc2a})--(\ref{Biased-Hc2b});
    dotted lines describe the formation of edge-assisted parallel-domain superconductivity (EA-PDS);
    while black dash-dotted line describes the appearance of domain-wall superconductivity (DWS),
    Eq.~(\ref{Aladyshkin-DWS}).
    Here we used the following fitting parameters:
    $T_{c0}=7.25~$K, $B_0=480~$Oe, and $H_{c2}^{(0)}=2.25~$kOe.
    The black dots correspond to the low-temperature scanning laser microscopy
    images obtained at $T=4.60$~K and $T=5.70$~K
    and which are presented in Figs.~\ref{Fig-LTSLM-1}--\ref{Fig-LTSLM-2}, respectively. \newline
    (c) Schematic presentation of the different regimes of inhomogeneous superconductivity in the considered system.
    Bright and dark areas correspond to positive and negative magnetic domains respectively,
    shaded areas depict the expected superconducting regions:
    I and II -- EA-RDS above positive and negative magnetic domains,
    respectively;
    III -- complex state consisting of DWS and EA-RDS;
    IV and V -- bulk RDS above positive and negative domains,
    respectively;
    VI and VII -- complex states consisting of RDS and EA-PDS above positive and negative domains,
    respectively;
    VIII -- complete superconductivity in the entire sample, consisting of RDS and PDS.
    }
    \label{Fig-Experiment-PB}
    \end{figure*}

Typical $R(H)$ curves measured at $I=100~\mu$A are shown in
Fig.~\ref{Fig-resistance}(a). The appearance of two symmetrical
minima in $R(H)$ at $T=7.20$~K corresponds to RDS above the
domains of opposite polarity. Taking the positions of the $R$
minima, one can estimate the amplitude of the nonuniform magnetic
field $B_0$ inside the superconducting bridge to 480~Oe. The
observed linear increase in the width of the $R$ minima with
decreasing temperature (compare the curves from $T=7.20$~K, and to
$6.50$~K) allows us to prove the usual relationship
$H_{c2}=H_{c2}^{(0)}\,(1-T/T_{c0})$ (Eq.~\ref{Hc2}) and to
estimate both the maximal critical temperature $T_{c0}=7.25$~K and
\mbox{$H_{c2}^{(0)}\simeq 2.25\times 10^3$~Oe}. Contrary to the
previously studied hybrid Al/BaFe$_{12}$O$_{19}$
bilayers,\cite{Aladyshkin-APL-09,Aladyshkin-PhysC-10,Aladyshkin-APL-10}
the estimated $H_{c2}^{(0)}$ value for Pb is substantially higher
than $B_0$ at low temperatures. The ratio $B_0/H_{c2}^{(0)}\simeq
0.2$ gives us the temperature of the anticipated crossover
$T^*=5.70$~K between RDS and CS. The $R(H)$ at $T=5.70$~K confirms
this simple estimate since the sample resistance vanishes for
$T<5.70$~K at $H=0$ which indicates CS.

Considering the $R(H)$ curves at rather low temperatures
($T<T^*$), one can see that the transition from the
superconducting to the normal state upon increasing $|H|$ from
zero occurs in two stages. The first stage and an appearance of
nonzero resistance can be attributed to the suppression of bulk
PDS above the parallel domains since the position of this anomaly
corresponds to Eqs.~(\ref{Biased-Hc2a})--(\ref{Biased-Hc2b}).
However, due to the presence of a continuous superconducting path
along the sample edges attributed to edge-assisted
superconductivity (PDS), the resistance increases slowly as $H$
increases. The exclusive survival of RDS above opposite domains at
larger $H$ explains the rise in the total sample's resistance $R$
up to 50-60\% from the normal state resistance $R_n$ (curves
4.50~K, 4.90~K, 5.30~K and 5.70~K). Indeed, for the considered
topology of the magnetic field, the superconducting and normal
regions are connected in series and, therefore, $R$ should reflect
the ratio between the volume of the bridge in the normal state and
the total volume of the bridge. We observe that in the developed
RDS state $R$ is not constant but slightly increases with $|H|$.
This finding can be attributed to the external-field-induced
shrinkage of the reverse domains. However, even when bulk RDS is
suppressed, the resistance is still lower than the normal
resistance. Such a resistive state observed at low temperature and
high field, can be attributed to the formation of compensated
superconductivity above magnetic domains of opposite polarity but
localized near the edge of the sample. In the following, this
state will be referred to as edge-assisted RDS. Apparently, such
states can exist until the local magnetic field above the opposite
domains $|H-B_0|$ exceeds the critical field of surface
superconductivity $H_{c3}=1.69 H_{c2}$ at a given
temperature.\cite{Saint-James-69,Tinkham-96} The position of the
critical fields $|H_{c2}+B_0|$ and $|H_{c3}+B_0|$, which determine
the shape of the magnetoresistive curve, are shown in
Fig.~\ref{Fig-resistance}(b) for $T=6.10$~K. For instance, the
experimentally determined ratio $H_{c3}/H_{c2}$ for $T=6.10$~K is
close to 1.77, supporting our interpretation.

\vspace{0.2cm}

    \begin{figure*}[htb!]
    \ifgraph\includegraphics[width=18.2cm]{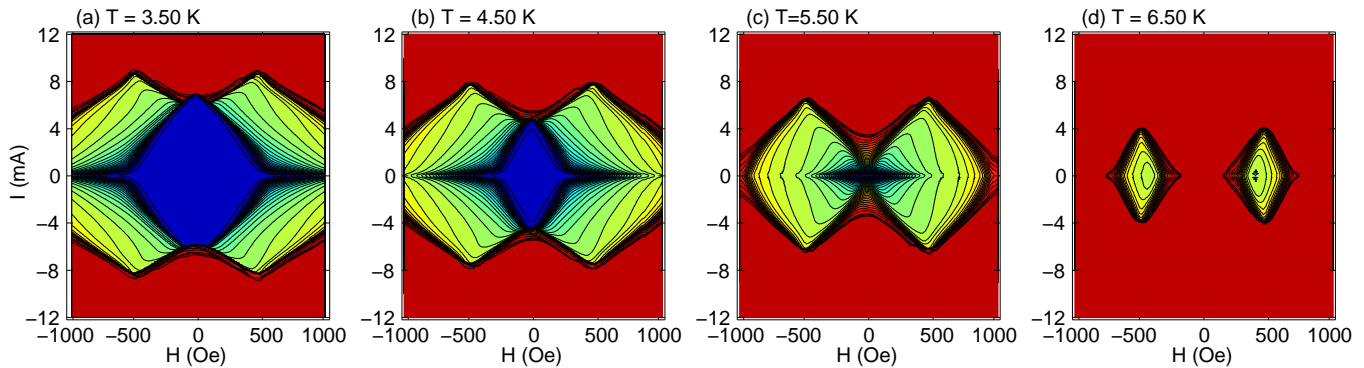}\fi
    \caption{(Color online) Dc resistance $R=V/I$ dependence of the
    long superconducting Pb bridge on the external magnetic
    field $H$ and the bias current $I$ taken at different
    temperatures: $T=3.50$~K (a), $T=4.50$~K (b), $T=5.50$~K (c), and
    $T=6.50$~K (d). We use the same color scheme as in Fig.~\ref{Fig-Experiment-PB}(a).}
    \label{Fig-2D-Currents}
    \end{figure*}

{\it $H-T$ diagram.} A full $H-T$ diagram for the long bridge,
composed from isothermal $R(H)$ measurements, is presented in
Fig.~\ref{Fig-Experiment-PB}(a). A similar diagram for the short
bridge is not given since both S/F hybrid samples showed almost
identical behavior. The interpretation of all distinctive regions
of this diagram is given in Fig.~\ref{Fig-Experiment-PB}(b).

We explain the initial deviation of the resistance from the normal
value $R_n$ by the formation of the edge-assisted RDS for large
$H$. Such states I and II are shown schematically in
Fig.~\ref{Fig-Experiment-PB}(c). For moderate $H$,
superconductivity can appear in the form of a complex state
consisting of edge-assisted superconductivity and DWS (pattern
III). According to our expectations, the localized
superconductivity in the latter case ($|H|<B_0$) appears above the
phase transition line depicted by Eq.~(\ref{Aladyshkin-DWS}),
since the critical temperature for DWS induced by the domain walls
of a finite width should be always higher than that for infinitely
narrow domain walls. Interestingly, in the overcompensated regime
($|H|>B_0$), the position of the shifted $H_{c3}$ line,
Eq.~(\ref{Biased-Hc3}), coincides with the level curve
$R(H,T)=0.98\,R_n$. The next stages of decreasing resistance with
decreasing temperature at $|H|\sim B_0$ have to be associated with
the appearance of bulk RDS in the compensated regions (patterns IV
and V). The corresponding phase boundaries are given by
Eqs.~(\ref{Biased-Hc2a})--(\ref{Biased-Hc2b}) and are shown by
black dashed lines in Fig.~\ref{Fig-Experiment-PB}(b). For
temperatures  below than the transition line given by
Eq.~(\ref{Biased-Hc3}), in addition to RDS inhomogeneous
superconductivity in the form of edge-assisted PDS also appears in
the regions with enhanced magnetic field (patterns VI and VII).
Inside the region VIII in the $H-T$ diagram, where areas of
inhomogeneous superconductivity for the different domains start to
overlap, the total absence of electrical resistance indicates the
appearance of CS.

Thus, in our transport measurements, we clearly observed the
switching between different regimes of localized superconductivity
upon variation of $H$ and $T$.

    \begin{figure}[htb!]
    \includegraphics[width=8.2cm]{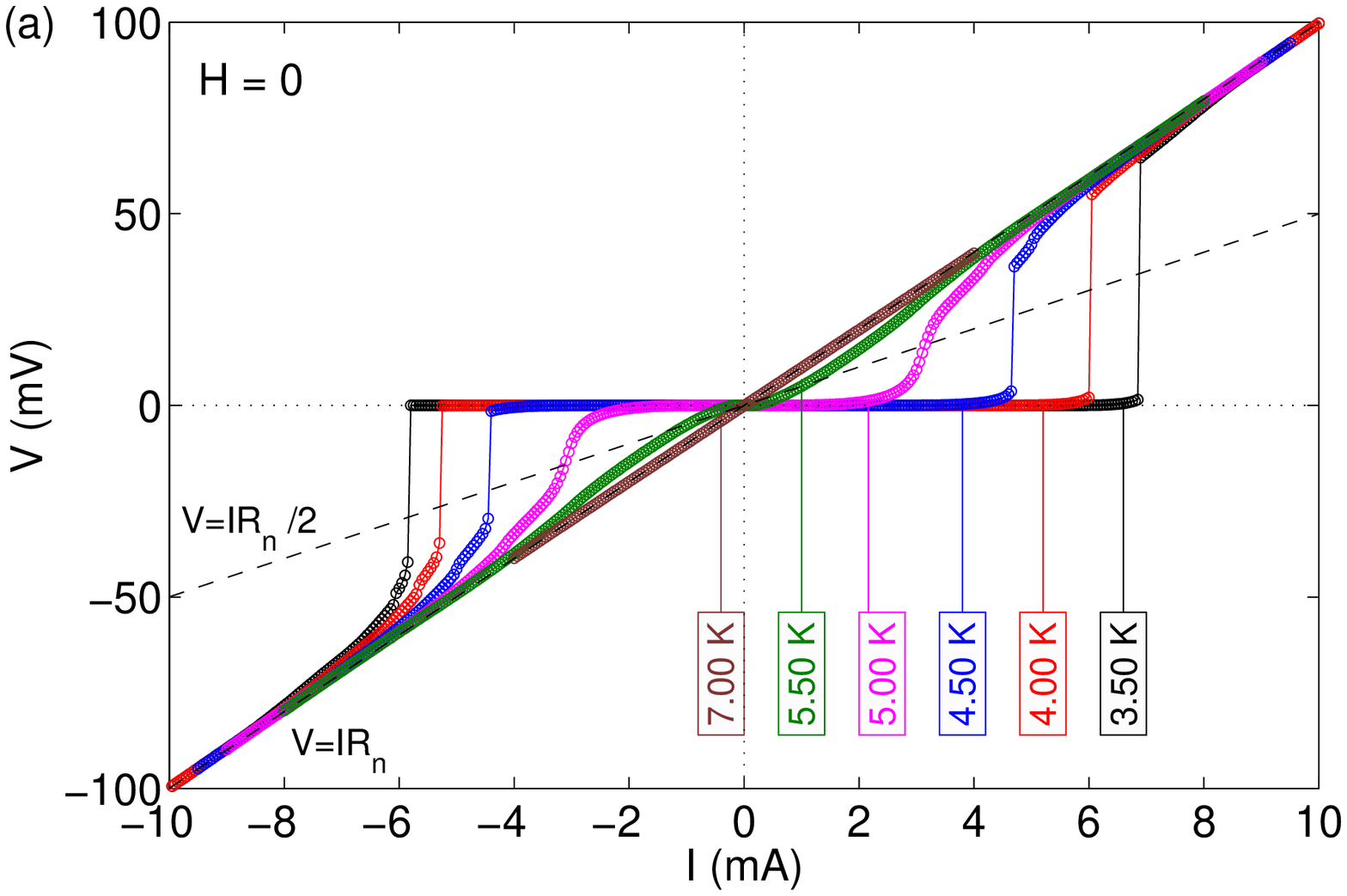}
    \includegraphics[width=8.2cm]{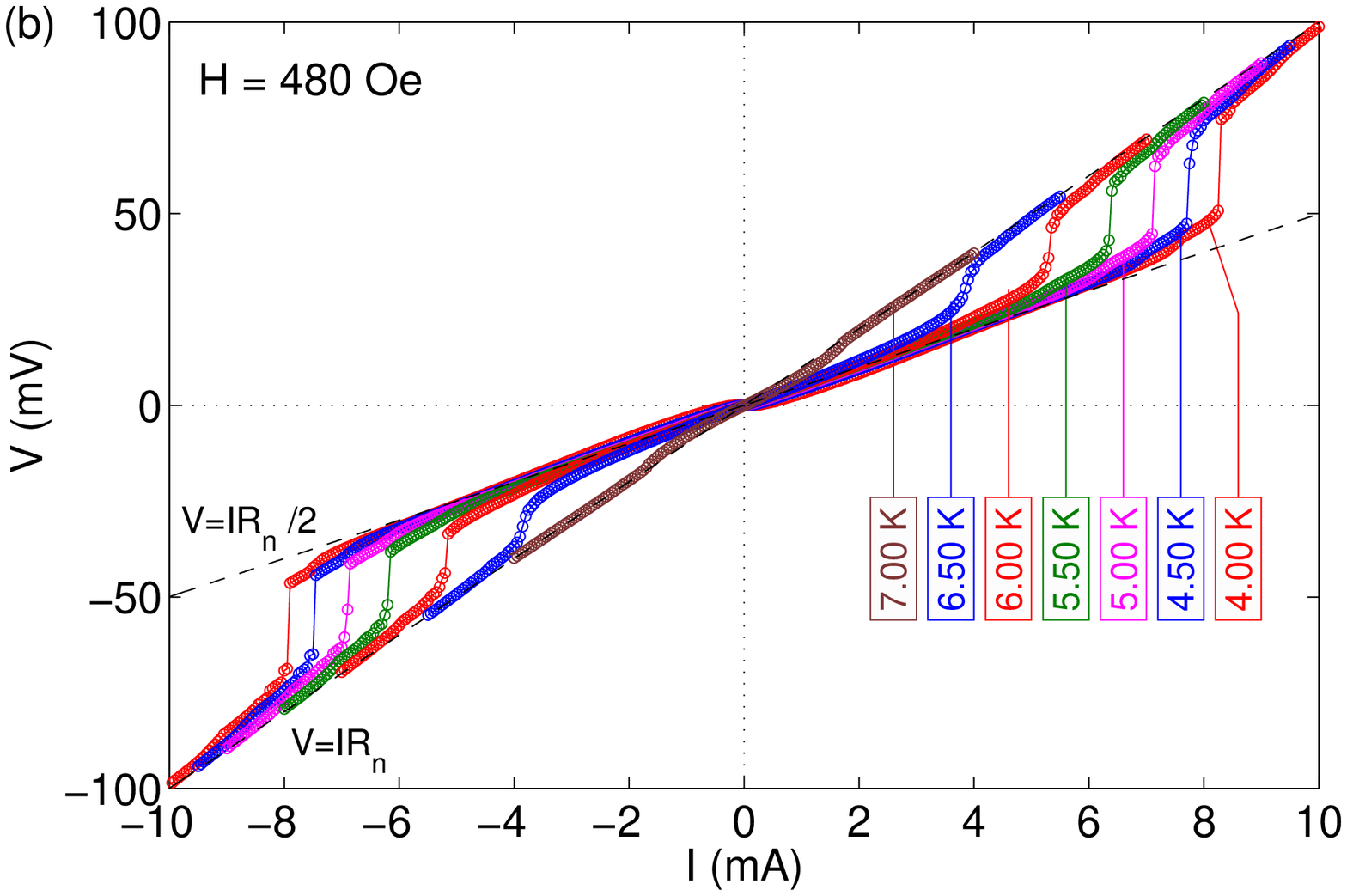}
    \caption{(Color online) Typical $I-V$ dependencies measured at $H=0$ (a) and $H=480$~Oe (b)
    for the long superconducting Pb bridge at different temperatures indicated in the plots. Two dashed lines
    correspond to the Ohmic-like behavior with the resistance $R=R_n/2$ and $R=R_n$.}
    \label{Fig-IV}
    \end{figure}

   \begin{figure}[htb!]
    \ifgraph\includegraphics[width=8.4cm]{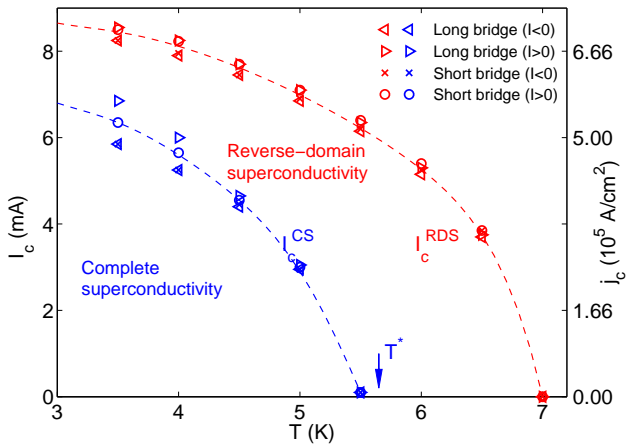}\fi
    \caption{(Color online) Temperature dependence of the critical
    currents $I_c^{RDS}$ and $I_c^{CS}$, corresponding to the suppression of
    reverse-domain superconductivity (red symbols near the top line), and
    complete superconductivity (blue squares near the bottom line), respectively.
    For the determination of $I_c^{RDS}$ ($I^{CS}$) we plotted the positions of the maximal differential resistance on
    the $I-V$ dependencies measured at
    $H=480~$Oe ($H=0$) both for the long and short bridges; $I<0$
    and $I>0$ denote the negative and positive branches of the
    corresponding $I-V$ curves (Fig.~\ref{Fig-IV}).
    The dashed lines are guides to the eyes.} \label{Fig-CritCurrents}
    \end{figure}

\vspace{0.2cm}

{\it Critical currents.} A bias current $I$ can suppress the
different modes of nonuniform superconductivity in the considered
S/F system in various ways. The effect of $H$ and $I$ on the
current ($I$) -- voltage ($V$) dependencies and on the dc
resistance $R=V/I$ are illustrated in Figs.~\ref{Fig-2D-Currents}
and \ref{Fig-IV}. We observed that the state of RDS is more robust
with respect to current injection than the state of CS, since in
the latter case the bias current can first destroy the
weakly-developed superconductivity above parallel magnetic domains
(i.e., above positive domains at $H>0$ and vice versa), where the
local magnetic field is maximal.

In order to evaluate the critical current destroying CS and RDS,
we consider the effect of $T$ on the $I-V$ dependencies
(Fig.~\ref{Fig-IV}) measured at $H=0$ and $|H|\simeq B_0$ (i.e.
close to the compensation field). One can see that the state with
zero resistance, which is inherent to CS, can be destroyed if $I$
exceeds a threshold value $I_c^{CS}$. Taking the position of the
jumps in the $I-V$ curves at $H=0$ or the maximal differential
resistance $dR/dH$ (panel (a) in Fig.~\ref{Fig-IV}), we plotted
the the temperature dependence of the critical current $I_c^{CS}$,
corresponding to the destruction of the most developed CS state at
$H=0$. For a characterization of the critical current $I_c^{RDS}$,
corresponding to the suppression of the developed RDS at the
compensation field ($|H|\simeq B_0$), we traced the position of
the jump on the $I-V$ curves at $H=480~$Oe upon increasing $T$
(panel (a) in Fig.~\ref{Fig-IV}). Indeed, this estimate seems to
be reasonable for the describing of the current-induced
destruction of inhomogeneous superconductivity under the condition
when the minimal resistance of the investigated sample in the
superconducting state is finite (of the order of 50\%$R_n$). The
dependencies of the both estimated critical currents $I_c^{RDS}$
and $I_c^{RDS}$ as a function of $T$ are presented in
Fig.~\ref{Fig-CritCurrents}.

%


\section{Visualization of nonuniform superconducting states by scanning laser microscopy}

In order to image the inhomogeneous
superconducting states trapped
by the nonuniform magnetic field we used low-temperature scanning
laser microscopy (LTSLM).

    \begin{figure*}[hbt!]
    \begin{center}
    \includegraphics[width=13.5cm]{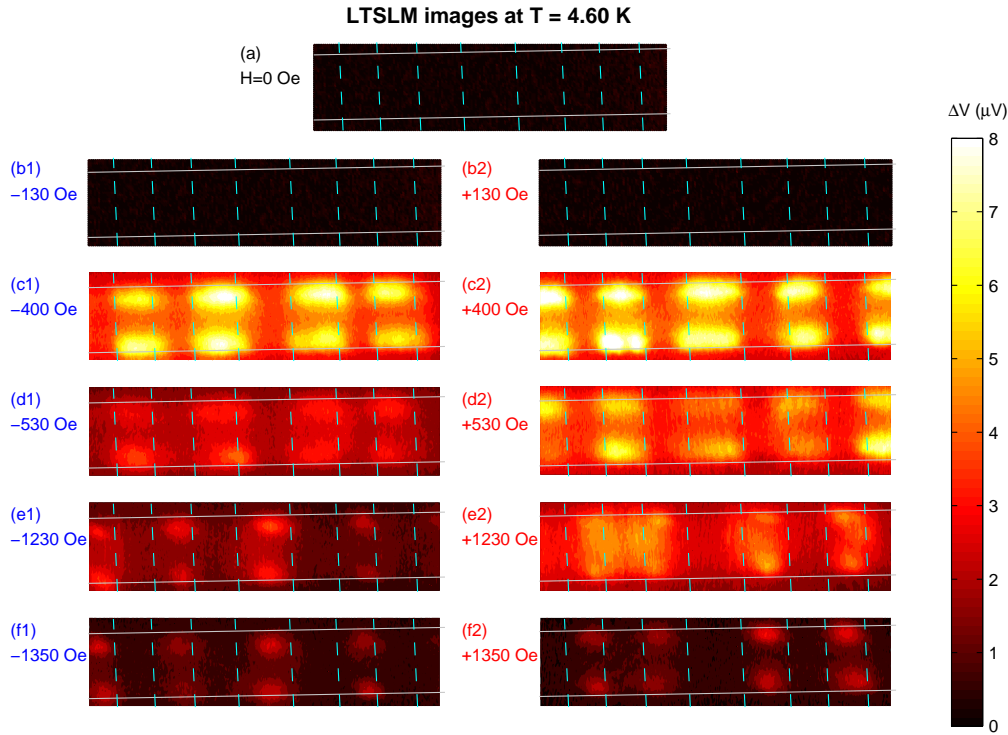}
    \end{center}
    \caption{(color online) Low-temperature scanning laser microscopy (LTSLM) images obtained
    for the same area of the long Pb bridge at
    \mbox{$T=4.60$\,K}. The color scale indicates the change in the beam-induced voltage drop $\Delta V$.
    The $H$ values are indicated in the plots and the bias points for all presented images are also marked by black dots in
    the phase diagram in Fig.~\ref{Fig-Experiment-PB}(b).
    The scanning area is about 120$\,\mu$m $\times$ 40$\,\mu$m.
    Vertical dashed cyan lines indicate the positions of the domain
    walls. The sample edges are marked by horizontal solid white lines. }
    \label{Fig-LTSLM-1}
    \end{figure*}

The principle of operation of LTSLM can be introduced as
follows.\cite{Fritzsche-PRL-06,Testardi-PRB-71,Fritzsche-Thesis,Wagenknecht-PRL-06,Wang-PRL-09}
The hybrid S/F samples were mounted on the cold finger of a Helium
gas flow cryostat, which is equipped with optical windows to
enable laser irradiation. An amplitude modulated laser beam (wave
length 680 nm, modulation frequency 10~kHz) heats locally the
superconducting sample within a spot with a diameter of
$1.5-2\,\mu$m. This value is determined by the diameter of the
focused incident beam and by the thermal conductivity of the
tested bilayer sample.\cite{Fritzsche-Thesis} The incident beam
intensity power on the sample surface (up to $\sim 25~\mu$W)
appears to be high enough to provide a maximum beam-induced
increase in temperature $\Delta T\sim 0.1-0.2~$K, leading to a
local suppression of superconductivity within this hot spot. We
assume that the effect of the laser irradiation should be uniform
across the superconducting film thickness since the thickness of
the Pb film (40 nm) is much smaller than the lateral spot size. In
our LTSLM measurements we apply a constant bias current and
measure the beam-induced voltage drop $\Delta V$ along the entire
bridge by lock-in technique. A set of scanning mirrors allows us
to control the position $(x,y)$ of the spot and thus, to obtain
the position-dependent 2D map of the LTSLM signal: $\Delta
V=\Delta V(x,y)$.



The LTSLM voltage signal can be interpreted as follows. If the
laser spot heats an area of the bridge which is in the normal
resistive state, the beam-induced perturbation of the local
temperature causes only a very small change in the total
resistance, since $\partial R_n/\partial T$ is very small.
However, if the irradiated part of the bridge is in the
superconducting state and it took part in the transfer of a
substantial part of the supercurrents, the beam-induced
suppression of superconductivity might switch the whole sample
from a low-resistive state to a high-resistive state. In other
words, the LTSLM technique makes it possible to map out the
ability of the sample to carry supercurrents. Comparing the LTSLM
responses upon varying $H$ and $T$, one can trace the evolution of
local superconducting properties in the investigated system.


     \begin{figure*}[hbt!]
    \begin{center}
    \includegraphics[width=13.5cm]{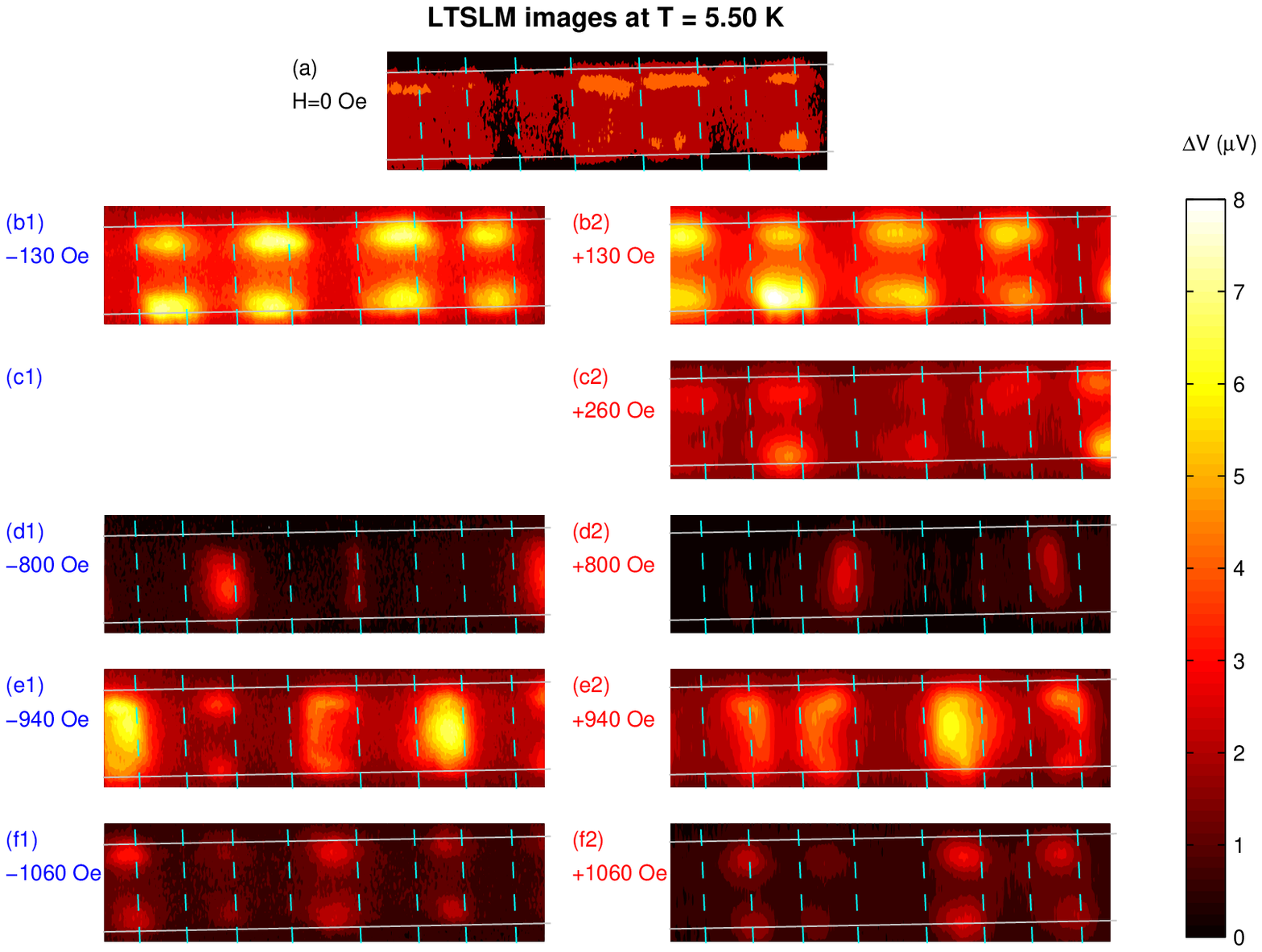}
    \end{center}
    \caption{(color online) Low-temperature scanning laser microscopy (LTSLM) images obtained
    for the same area of the long Pb bridge at
    \mbox{$T=5.50$\,K}. All parameters and notations are the same as in Fig.~\ref{Fig-LTSLM-1}.
    The LTSLM image at $H=-260~$Oe (panel (c1)) was recorded with
    technical problems and therefore we do not present this plot.}
    \label{Fig-LTSLM-2}
    \end{figure*}

For the observation of the different regimes of inhomogeneous
superconductivity in the long Pb/BFO hybrid samples, we applied a
constant current of $I=300~\mu$A and the field was varied in the
range between $H=\pm 1350$~Oe.

Figure~\ref{Fig-LTSLM-1} shows the LTSLM images obtained at
$T=4.60$\,K, which is below the crossover temperature
$T^*=5.70$~K. For $|H|\le 130~$Oe the measured responses have no
detectable variations [see panels (a), (b1) and (b2)]. Apparently,
at these points inside the CS state in the $H-T$ diagram, the
intensity of the laser beam is insufficient for the destruction of
the developed bulk superconductivity.

According to our estimates, at $T=4.60$\,K, the depletion of
superconductivity above the parallel magnetic domains should occur
at $|H|\gapprox$ 350~Oe. This means that near the ``CS--RDS"
transition line the areas above the parallel domains have to be in
the {\it resistive} state, while the areas above the opposite
domains are still in the superconducting state. As a result, the
maxima in the LTSLM response at $H=400$ Oe should be attributed
solely to the non-superconducting regions above the positive
domains [panels (c2), (d2)]. Correspondingly, at $H=-400$ Oe the
resistive areas above negative domains are responsible for the
beam-induced voltage [panels (c1), (d1)]. At higher $H$ values
(i.e. deeper in the RDS areas in the $H-T$ diagram)
superconductivity survives exclusively above the opposite domains
and such compensated superconductivity is strong enough not to be
destroyed by the laser beam of the given intensity. However, close
to the transition line ``RDS -- edge-assisted RDS" the
reverse-domain superconductivity becomes weaker and can be
affected by the laser beam. Therefore, the bright areas in the
corresponding LTSLM signal originate from the regions with the
compensated magnetic field above the opposite domains. The
inversion of the $H$ sign immediately results in a switching
between enhanced reverse-domain superconductivity and depleted
parallel-domain superconductivity for the same areas of the
superconducting bridge [panels (e1--e2) and (f1--f2)]. Thus, all
findings concerning the migration of the maximal beam-induced
voltage along the Pb bridge upon varying $H$ are in agreement with
our transport measurements.

We would like to note that the LTSLM images [panels (c1--f1) and
(c2--f2) in Fig.~\ref{Fig-LTSLM-1}] reveal an inhomogeneity of the
beam-induced voltage across the width of the bridge. Indeed, the
$\Delta V$ maxima are always located near the edges of the bridge.
At the present moment we have no reliable interpretation of this
effect. Probably, it can be explained by the current enhancement
near the sample edges, typical for plain superconducting
bridges,\cite{Kupriyanov-1974,Elistratov-2002} or by a suppression
of the energy barrier for flux entry or exit for superconductors
in the developed mixed state. Another possible explanation for the
pronounced edge signal is the manifestation of edge-assisted PDS,
since such edge-assisted PDS states seem to be the states with
weakest superconductivity. Finally, the edge signal can be
explained by a hampered heat diffusion and, correspondingly, a
larger heating effect of the beam focused near the edges as
compared to that in the interior of the bridge. In any case, this
issue requires a detailed theoretical treatment which is beyond
the scope of the present work.

At $T=5.50$~K, close to the crossover temperature, a voltage
signal can be detected at $H=0$. The amplitude of the built-in
magnetic field is close to the corresponding upper critical field
and therefore even a weak optical influence can substantially
suppress bulk superconductivity equally above the domains of both
polarities. Upon increasing $|H|$ we successively observe the
responses from the parallel domains [panels (b1) and (b2-c2)] and
from the antiparallel domains [panels (d1-f1) and (d2-f2)],
similar to that described above.

    \begin{figure}[t!]
    \begin{center}
    \includegraphics[width=8.5cm]{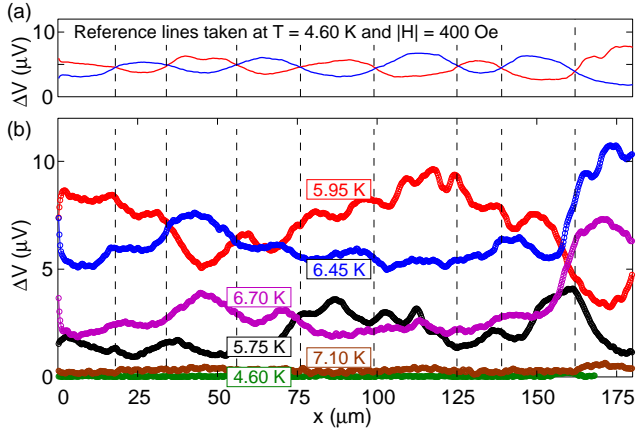}
    \end{center}
    \caption{(color online) Line scans $\Delta V(x)$ obtained by LTSLM method at different $T$
    along the line close to the lower edge of
    the Pb microbridge: (a) the reference signal $\Delta V(x)$
    measured at $H=\pm 400~$Oe in order
    to detect the positions of the domain walls shown as vertical dashed lines. (b) The dependencies
    $\Delta V(x)$ measured at different temperatures (shown in the plot) and $H=0$.}
    \label{Fig-LTSLM-3}
    \end{figure}

Further increase in temperature above the crossover temperature
could allow us to detect the domain-wall superconductivity at
$H=0$, since the masking background signal from CS and the
edge-assisted RDS are turned off. The evolution of the
beam-induced voltage upon increasing $T$ is presented in
Fig.~\ref{Fig-LTSLM-3}, where we show line scans $\Delta V(x)$
along the bridge, close to the bottom edge of the bridge. We did
not find any noticeable increase in the LTSLM response near the
domain walls for the considered S/F system. We would like to
mention that the DWS state was observed for a similar Pb bridge
oriented along the domain wall using the same LTSLM
technique.\cite{Werner-2011}



\section{Conclusion}

We presented a detailed experimental study of the superconducting
properties of thin-film superconducting Pb microbridges in the
presence of a nonuniform magnetic field of the laminar domain
structure in ferromagnetic BaFe$_{12}$O$_{19}$ crystals. Such
ferromagnets generate rather strong stray fields and the
parameters of this field (amplitude and period) are almost
independent on the applied magnetic field $H$ in the considered
$H$ range. We focused on the case when the domain walls are
oriented perpendicular to the bridge in order to avoid electrical
shunting and masking of less developed superconducting states by
more favorable states during transport measurements. It was
demonstrated that, at high temperatures superconductivity appears
in the form of reverse-domain superconductivity only above
magnetic domains of opposite polarity with respect to the $H$
sign. Below the crossover temperature $T^*$, defined as
$H_{c2}(T^*)=B_0$, superconductivity can nucleate both above
antiparallel and parallel magnetic domains ($B_0$ is the amplitude
of the perpendicular component of the nonuniform field,
\mbox{$H_{c2}=H_{c2}^{(0)}\,(1-T/T_{c0})$} is the
temperature-dependent upper critical field). Indeed, at $T<T^*$
the regions of inhomogeneous superconductivity above positive and
negative domains overlap in the $H-T$ plane, resulting in
so-called complete (or global) superconductivity and in the
vanishing total electrical resistance of the hybrid sample. We
also found experimental evidences for the regimes of edge-assisted
reverse-domain superconductivity and edge-assisted parallel-domain
superconductivity, corresponding to localized superconductivity
near the edges of the bridge above the regions with compensated
and enhanced magnetic field, respectively. We experimentally
determined the critical currents, corresponding to the suppression
of the localized superconductivity above parallel and antiparallel
magnetic domains in a broad temperature range. The technique of
low-temperature scanning laser microscopy made it possible to
directly visualize the temperature-- and field-induced transitions
from complete superconductivity to reverse-domain
superconductivity and parallel-domain superconductivity and from
these inhomogeneous superconducting states to the normal state.

\section{Acknowledgements}

This work was supported by the Methusalem Funding of the Flemish
Government, the NES -- ESF program, the Belgian IAP, the Fund for
Scientific Research -- Flanders (F.W.O.--Vlaanderen), the Russian
Fund for Basic Research, RAS under the Program ``Quantum physics
of condensed matter", Russian Agency of Education under the
Federal Target Program ``Scientific and educational personnel of
innovative Russia in 2009--2013", Deutsche Forschungsgemeinschaft
(DFG) via grant no. KO 1303/8-1. Robert Werner acknowledges
support by the Cusanuswerk, Bisch\"{o}fliche Studienf\"{o}rderung.


\begin{references}
\bibitem{Comment-2}
Equation~(\ref{Aladyshkin-DWS}) may be considered as an ultimate
estimate of DWS, since the critical temperature of the solution,
localized in vicinity of the domain walls of the finite width,
will be always higher than that given by
Eq.~(\ref{Aladyshkin-DWS}).
\end{references}

\end{document}